\documentclass[10pt]{article}
\PassOptionsToPackage{dvipsnames,table}{xcolor}
\PassOptionsToPackage{most}{tcolorbox}
\PassOptionsToPackage{numbers,sort&compress}{natbib}
\usepackage[wide]{preprint}
\setcitestyle{numbers,square,comma}
\usepackage[utf8]{inputenc}
\usepackage{microtype}
\usepackage{latexsym}
\usepackage{pifont}
\usepackage{xspace}
\usepackage{amsthm}
\usepackage{fontawesome5}
\usepackage{wrapfig}
\usepackage{subcaption}
\usepackage{multirow}
\usepackage{algorithm}
\usepackage{algorithmic}
\usepackage{makecell}
\usepackage{bbm}
\usepackage{hhline}
\usepackage{placeins}
\usepackage{tablefootnote}
\usepackage{adjustbox}
\usepackage{hyperref}

\usepackage{paper-layout}

\def\eg{\emph{e.g., }\xspace}

\definecolor{rqlocalrow}{RGB}{247,250,255}
\definecolor{rqapirow}{RGB}{247,252,247}
\definecolor{rqavgrow}{RGB}{246,246,246}

\definecolor{deepred}{RGB}{139,0,0}
\definecolor{deepgreen}{RGB}{0,100,0}
\definecolor{tagblue}{RGB}{52, 101, 175}

\def\attack{\textcolor{deepred}{\textit{CodeSpear}}\xspace}
\def\defense{\textcolor{deepgreen}{\textit{CodeShield}}\xspace}
\def\attackt{\textcolor{deepred}{\textit{CodeSpear\textbf{-T}}}\xspace}

\def\qwencoderseven{Qwen2.5-Coder-7B\xspace}

\def\qwencoderthirtytwo{Qwen2.5-Coder-32B\xspace}
\def\qwenseven{Qwen2.5-7B\xspace}

\def\qwenthirtytwo{Qwen2.5-32B\xspace}  
\def\llamaeight{LLaMA3-8B\xspace}

\def\gpt{GPT-5\xspace}
\def\gptmini{GPT-5-mini\xspace}
\def\deepseekpro{DeepSeek-V4-Pro\xspace}
\def\minimaxnew{MiniMax-M2.7\xspace}
\def\minimaxold{MiniMax-M2.5\xspace}
\def\oss{GPT-OSS-120B\xspace}

\newcommand{\bi}[1]{\textbf{\textit{#1}}}
\newcommand{\pass}[1]{\textit{pass@#1}}

\newcommand{\baselinename}[1]{\textit{#1}\xspace}
\newcommand{\dataname}[1]{{\fontfamily{cmtt}\selectfont {#1}}\xspace}

\definecolor{namerow}{RGB}{246,249,255}

\definecolor{vanillarow}{RGB}{252,252,252} 
\definecolor{ragrow}{RGB}{246,252,255}     
\definecolor{newnamerow}{RGB}{246,253,246} 

\newtcolorbox{boxK}{
  colback=tagblue!5, colframe=tagblue!40, boxrule=0.5pt, arc=3pt,
  left=6pt, right=6pt, top=4pt, bottom=4pt
}

\newcommand{\summary}[2]{%
\begin{boxK}
\small \ding{46} \textbf{#1:}
\textit{#2}
\end{boxK}
}

\definecolor{lightred}{RGB}{252,232,232}
\definecolor{darkred}{RGB}{150,45,45}

\newcommand{\redcircle}[1]{%
  \tikz[baseline=-0.75ex]{
    \node[
      shape=circle,
      fill=lightred,
      text=darkred,
      inner sep=1.2pt,
      minimum size=1.15em,
      font=\scriptsize\bfseries
    ] {#1};
  }%
}

\definecolor{lightgreen}{RGB}{220,245,225}
\definecolor{darkgreen}{RGB}{35,120,70}

\newcommand{\greencircle}[1]{%
  \tikz[baseline=-0.75ex]{
    \node[
      shape=circle,
      fill=lightgreen,
      text=darkgreen,
      inner sep=1.2pt,
      minimum size=1.15em,
      font=\scriptsize\bfseries
    ] {#1};
  }%
}

\def\BibTeX{{\rm B\kern-.05em{\sc i\kern-.025em b}\kern-.08em
    T\kern-.1667em\lower.7ex\hbox{E}\kern-.125emX}}

\logos{%
  \mbox{%
    \raisebox{-0.5\height}{%
      \includegraphics[height=0.30in,
        trim=465bp 80bp 490bp 75bp,clip]{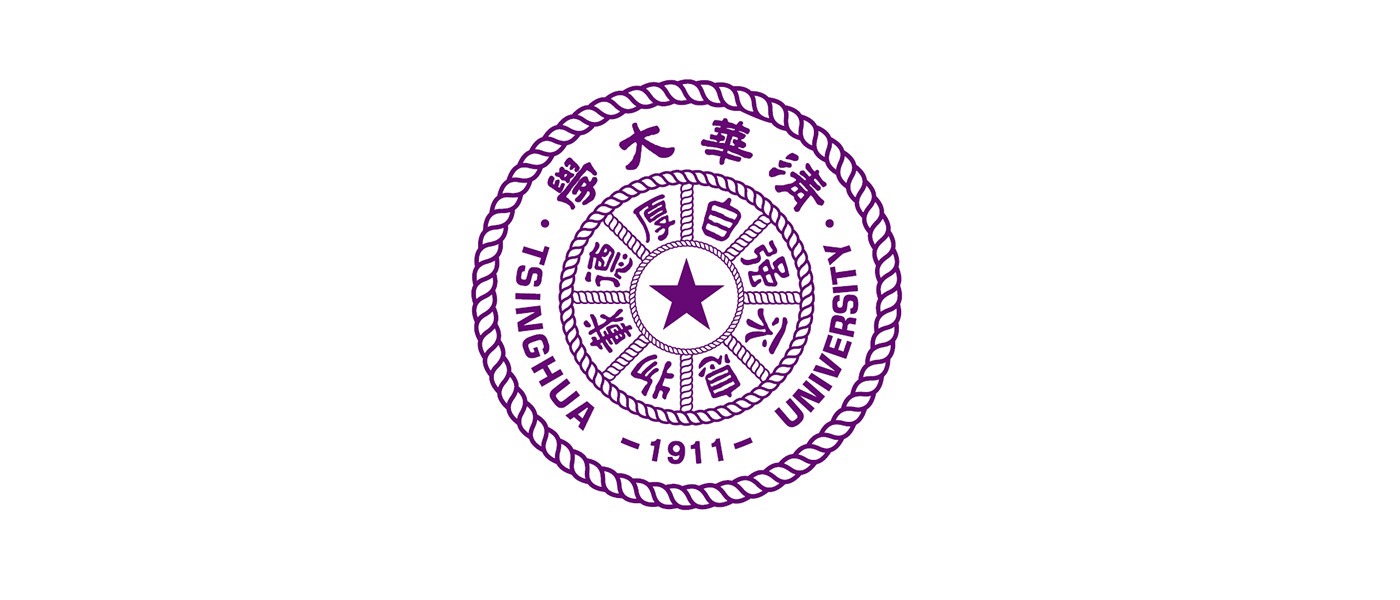}}%
    \hspace{7pt}%
    \raisebox{-0.5\height}{%
      \includegraphics[height=0.32in,
        trim=0bp 0bp 0bp 120bp,clip]{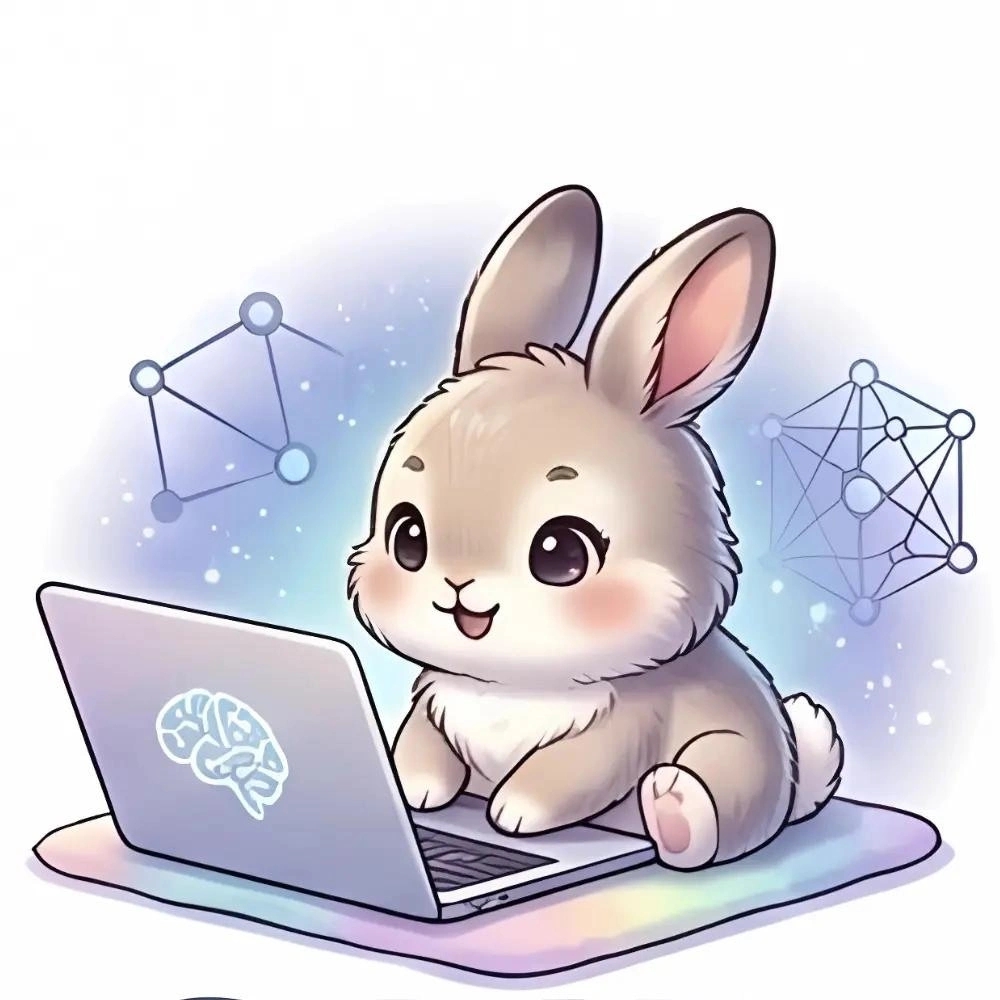}}%
    \hspace{7pt}%
    \raisebox{-0.5\height}{%
      \fontfamily{Montserrat-LF}\fontseries{medium}\fontshape{n}%
      \fontsize{10.5}{12}\selectfont\color{accent}%
      \textls[10]{\groupname}}%
  }%
}

\title{Grammar-Constrained Decoding Can Jailbreak LLMs into Generating Malicious Code}
\author[1,*]{Yitong Zhang}
\author[2,*,\ddagger]{Shiteng Lu}
\author[1,\dagger]{Jia Li}
\affiliation{%
  $^{1}$College of AI, Tsinghua University, Beijing, China\\
  $^{2}$Yingcai Honors College, University of Electronic Science and Technology of China, Chengdu, China}
\metadata[Yitong Zhang]{\email{zhangyt26@mails.tsinghua.edu.cn}}
\metadata[Shiteng Lu]{\email{lushiteng@std.uestc.edu.cn}}
\metadata[Corresponding author]{$^{\dagger}$\,\email{jia\_li@mail.tsinghua.edu.cn}}

\makeatletter
\newcommand{\paperauthornotes}{%
  \begingroup
  \renewcommand{\thefootnote}{}%
  \long\def\@makefntext##1{\noindent##1}%
  \footnotetext[0]{%
    \begin{tabularx}{\linewidth}{@{}p{0.9em}@{}>{\raggedright\arraybackslash}X@{}}
      {*} & Equal contribution: Yitong Zhang proposed the idea and wrote the paper; Shiteng Lu implemented the approaches and ran most of the experiments.\\
      \textsuperscript{\ddag} & This work was done while Shiteng Lu was an intern at the College of AI, Tsinghua University.\\
      \textsuperscript{\dag} & Corresponding author.
    \end{tabularx}}%
  \endgroup
}
\makeatother

\begin{document}
\maketitle
\paperauthornotes

\begin{abstract}
Large Language Models (LLMs) are increasingly used for code generation, raising concerns that they may be misused to produce malicious code.
Meanwhile, Grammar-Constrained Decoding (GCD) has been widely adopted to improve the reliability of LLM-generated code by enforcing syntactic validity.
In this paper, we reveal a counterintuitive risk: this reliability-oriented technique can itself become an attack surface.
We uncover a new jailbreak attack, termed \attack, that exploits GCD to induce LLMs into generating malicious code.
Our experiments show that simply applying a benign code grammar constraint can effectively jailbreak LLMs.

To address this vulnerability, we propose \defense, a safety alignment approach that robustly preserves safe behavior even under attacker-controlled grammar constraints.
\defense aligns the model in the code modality by teaching it to generate \emph{honeypot code} under GCD.
Such code is semantically harmless, so it does not implement the malicious request, and structurally diverse, so it is difficult to suppress through grammar tightening.
At the same time, \defense still preserves natural-language refusals when natural language is available.
Experiments on 10 popular LLMs across 4 benchmarks show that \attack outperforms representative jailbreak baselines and increases the attack success rate by more than 30 percentage points on average over unconstrained decoding.
\defense also restores safety under \attack while preserving benign utility.
Our findings reveal a fundamental risk of GCD and call for greater attention to its potential security implications.
\end{abstract}


\section{Introduction}
\label{1-intro}

Large Language Models (LLMs) have demonstrated impressive capabilities across a wide range of tasks and are increasingly deployed in real-world applications~\cite{chen2025beyond, zhang2026see, cai2025ai}. At the same time, growing evidence shows that LLMs can be jailbroken to bypass safety alignment and produce harmful content~\cite{zhang2026davsp, li2025diffuguard, zhang2024jailbreak, wang2026omni}. This risk becomes especially concerning in code generation~\cite{ouyang2025smoke, guo2025redcodeagent, wahed2025mocha}, where harmful outputs are not merely textual instructions but executable programs that can be directly weaponized against digital systems~\cite{cheng2025security, liu2026packmonitor}.

In this paper, we uncover a new jailbreak attack, termed \textbf{\attack}, that leverages widely used \bi{Grammar-Constrained Decoding (GCD)}~\cite{dong2024xgrammar, ugare2024syncode, llguidance} to induce LLMs into generating malicious code at a low cost. GCD was originally designed to improve the reliability of code generation by constraining LLMs to produce outputs that conform to a target grammar~\cite{zhang2026lookahead, netz2024using}, and it is now supported by many mainstream inference frameworks such as \textsc{vLLM}~\cite{vllm} and \textsc{SGLang}~\cite{sglang}. However, we find that this benign reliability mechanism can unexpectedly become an attack surface. When an LLM is asked to generate malicious code, simply applying GCD with a standard code grammar can prevent the model from expressing its refusal behavior and cause it to produce malicious code, even though the grammar itself is entirely benign. Our evaluation shows that \attack can jailbreak 10 popular LLMs (\eg \gpt~\cite{gpt-5}, \minimaxnew~\cite{minimax2026m27}, and \qwencoderthirtytwo~\cite{qwencoder}) and increase the attack success rate by more than 30 percentage points on average over \baselinename{Vanilla}.

We attribute the success of \attack to a limitation of existing safety alignment~\cite{pku-saferlhf, qi2025safety}: it is almost \bi{exclusively grounded in the natural-language modality}. 
Existing safety alignment typically teaches LLMs to respond to malicious requests with natural-language refusals such as \textit{``I cannot assist with that''}, implicitly assuming that natural language remains available at inference time~\cite{mou2025decoupling, pku-saferlhf}. However, GCD breaks this assumption. Once a code grammar is enforced, natural-language refusals fall outside the valid output space, and the model can no longer express the refusal behavior it learned during alignment. The model is therefore forced to continue generation in the code modality, where it has not been explicitly aligned to behave safely. This explains why a benign code grammar can make \attack effective, and raises a key question: \bi{how should safety alignment be performed in the code modality when natural-language refusals are unavailable?}

To this end, we propose \textbf{\defense}, a safety alignment approach for the \bi{code modality} that trains the model to generate \bi{honeypot code} against GCD. Honeypot code is \emph{semantically harmless} and \emph{structurally diverse}: it does not implement the malicious request, and it shows diverse syntactic structures. This design directly addresses the two requirements of safe behavior in the code modality.
\ding{182} First, the response must remain harmless even when the model is forced to generate code. \ding{183} Second, the response must be hard to remove by grammar tightening. This second requirement means that safe behavior should not be bound to a fixed code template. For example, teaching the model to always generate a refusal comment or a \texttt{pass} statement may appear safe and natural, but such behavior is tied to a narrow syntactic pattern. An attacker can simply use a tightened grammar that forbids comments and \texttt{pass} statements to reopen the attack (see Section~\ref{sec:rq2-result}). By contrast, honeypot code gives the model many harmless ways to stay within the code modality, making the safe behavior difficult to suppress without also excluding many structures needed by malicious programs (see Section~\ref{sec:adaptive}).

To evaluate \attack and \defense, we conduct comprehensive experiments on 10 LLMs across 4 benchmarks. Our results show that: \ding{182} \attack effectively bypasses the safety alignment of locally deployed LLMs (\eg \qwencoderseven), achieving an average attack success rate of 81.82\%; \ding{183} \attack also generalizes to commercial API-based LLMs (\eg \gpt), increasing the attack success rate by more than 40 percentage points on average; \ding{184} \defense restores model safety under GCD, reducing the attack success rate to a level even lower than that observed without any attack; \ding{185} \defense preserves benign utility, causing only minor degradation on benign code generation benchmarks; and \ding{186} both \attack and \defense remain stable across a wide range of hyperparameter settings.

In summary, we make the following contributions:
\begin{itemize}
\item We propose \attack, a jailbreak attack that leverages GCD to push LLMs into generating malicious code.
\item We introduce code-modality safety alignment through \defense, which teaches LLMs to generate honeypot code when natural-language refusals are unavailable.
\item We conduct comprehensive experiments, showing that \attack can bypass the safety alignment of existing LLMs, while \defense effectively restores their safety.
\end{itemize}

\section{Background and Related Work}
\label{sec:2-relate}

\subsection{Grammar-Constrained Decoding}
\label{sec:2-1-gcd}

Mainstream LLMs generate outputs by recursively choosing the next token from a vocabulary~\cite{llama, qwen, qwencoder}. Formally, let $M$ be an LLM with vocabulary $\mathcal{V}$. Given a prompt $p$, the model produces a response $y = (y_1, \ldots, y_T)$ token by token:
\begin{equation}
y_t \sim P_M(\cdot \mid p, y_{<t}), \quad y_{<t} = (y_1, \ldots, y_{t-1}).
\label{eq:ar-generation}
\end{equation}
In this standard decoding process, the model is free to produce any token sequence in $\mathcal{V}^{\ast}$. This freedom is essential for general open-ended dialogue, but it becomes a source of unreliability in code generation. Prior work~\cite{ugare2024syncode} has shown that even leading LLMs may assign nonzero probability to tokens that make the output syntactically invalid, causing the generated code to fail to parse, compile, or execute.

To mitigate this mismatch between the probabilistic generation of LLMs and the strict syntactic requirements of programming languages, Grammar-Constrained Decoding (GCD) has been introduced~\cite{ugare2024syncode, llguidance, dong2024xgrammar}.
Instead of allowing the model to sample from the full vocabulary, GCD typically restricts generation to a language defined by a grammar. 
Specifically, let $G$ be a code grammar and let $\mathcal{L}(G) \subseteq \mathcal{V}^{\ast}$ be the set of token sequences accepted by $G$. At each decoding step $t$, GCD efficiently computes the tokens that keep the current prefix extendable to some valid program~\cite{dong2024xgrammar}:
\begin{equation}
\mathcal{V}_G(y_{<t}) = \left\{\, v \in \mathcal{V} \;\middle|\; \exists\, y_{>t} \text{ s.t. } (y_{<t}, v, y_{>t}) \in \mathcal{L}(G) \,\right\},
\label{eq:valid-set}
\end{equation}
and masks all other invalid tokens by setting their probabilities to zero. Typically, the next token is then sampled from the renormalized distribution as follows:
\begin{equation}
P_M^{G}(y_t \mid p, y_{<t})
=
\frac{
P_M(y_t \mid p, y_{<t}) \cdot \mathbb{I}[y_t \in \mathcal{V}_G(y_{<t})]
}{
\sum_{v \in \mathcal{V}_G(y_{<t})} P_M(v \mid p, y_{<t})
}.
\label{eq:new-distribution}
\end{equation}
Thus, GCD induces the following output distribution:
\begin{equation}
P_M^{G}(y \mid p)
=
\mathbb{I}[y \in \mathcal{L}(G)]
\prod_{t=1}^{|y|}
P_M^{G}(y_t \mid p, y_{<t}),
\label{eq:gcd-response-distribution}
\end{equation}
where $P_M^{G}(y \mid p)=0$ for any sequence outside $\mathcal{L}(G)$. In effect, GCD leaves the prompt and model parameters untouched, but changes the support of the valid output space from $\mathcal{V}^{\ast}$ to $\mathcal{L}(G)$. 
Owing to its effectiveness, GCD is now natively supported by mainstream inference frameworks such as \textsc{vLLM}~\cite{vllm} and \textsc{SGLang}~\cite{sglang}, and is also exposed by popular commercial platforms including OpenAI~\cite{openai} and Fireworks AI~\cite{fireworks}.

This ability to reshape the output space also makes constrained decoding relevant to LLM safety. Existing studies~\cite{wang2026agentspec, liu2026packmonitor, zhou2026formaljudge} have primarily explored this technique from a defensive perspective, using constrained decoding to enforce safety-oriented rules. A smaller line of work~\cite{apt, enumattack} has shown that constrained decoding can also be used offensively to steer models toward unsafe content. However, such attacks typically depend on carefully crafted adversarial grammars tailored to specific malicious goals, which substantially limits their practicality and scalability. In contrast, our work reveals a more fundamental risk: even benign, off-the-shelf code grammars can be weaponized to induce malicious code generation.

\subsection{Jailbreaking and Safety Alignment of LLMs}
\label{sec:2-2-jailbreak}

Many jailbreak attacks have been proposed to expose the safety vulnerabilities of LLMs~\cite{dan, li2024lockpicking}. These attacks can be broadly categorized into three groups. The first group operates on the input by carefully crafting prompts that bypass safety measures~\cite{gcg-m, lrl, dan, swedeception}. For example, PAIR~\cite{pair} employs an attacker LLM to iteratively refine jailbreak prompts through feedback from the target model. The second group modifies the model itself. Prior studies~\cite{betley2025emergent, qi2024fine} have shown that fine-tuning on harmful data can weaken the safety alignment of target LLMs. The third group intervenes on the output side~\cite{apt, zhang2024jailbreak}. Rather than altering the prompt or the model weights, it manipulates the decoding process to induce unsafe outputs. For instance, JULI~\cite{wang2025juli} trains an auxiliary network to select unsafe tokens from the model's predicted logits. Our work falls into the third category, but differs from prior work in that it requires only a standard GCD interface invoked with a standard grammar, without any carefully crafted adversarial component.

Modern LLMs typically undergo safety alignment before deployment, which provides the main defense against potential misuse~\cite{pku-saferlhf, qi2025safety, safedpo}. Whether through supervised fine-tuning, preference optimization, or reinforcement learning over safety-related data, safety alignment aims to train the model to prefer safe responses on harmful prompts. Specifically, let $\mathcal{P}_{\text{mal}}$ be the set of malicious prompts and let $\mathcal{R}_{\text{refuse}} \subseteq \mathcal{V}^{\ast}$ be the set of refusal responses, such as \textit{``I cannot assist with that''} or \textit{``I am sorry.''} Existing safety alignment approaches typically aim to encourage $M$ to place most of its probability mass on $\mathcal{R}_{\text{refuse}}$ for any malicious prompt:
\begin{equation}
\Pr_{y \sim P_M(\cdot \mid p)}\!\left[\, y \in \mathcal{R}_{\text{refuse}} \,\right] \;\approx\; 1.
\label{eq:safety-alignment}
\end{equation}
However, this formulation implicitly grounds safe behavior in the natural-language modality, leaving safety in the code modality underexplored when natural language is unavailable.

Among existing alignment methods, Direct Preference Optimization (DPO)~\cite{dpo} is one of the most widely used and is the most relevant to our work. DPO is attractive because it directly optimizes pairwise response preferences without needing to train a separate reward model. Given a preferred response $y^{+}$ and a dispreferred response $y^{-}$ for a prompt $p$, DPO trains a model $M_\theta$ against a fixed reference model $M_{\text{ref}}$ by minimizing
\begin{equation}
\mathcal{L}_{\text{DPO}}
=
-\,
\mathbb{E}_{(p, y^{+}, y^{-})}
\left[
\log \sigma\!\left(
s(p, y^{+}) - s(p, y^{-})
\right)
\right],
\label{eq:dpo-generic}
\end{equation}
where $\sigma(\cdot)$ is the sigmoid function and the implicit reward is
$s(p, y) = \beta \log \frac{P_{M_\theta}(y \mid p)}{P_{M_{\text{ref}}}(y \mid p)}$,
with $\beta$ a temperature parameter. In standard safety alignment~\cite{safedpo}, these preference pairs usually take a simple form: the preferred response is a natural-language refusal, and the dispreferred response is a harmful completion. 
Our defense keeps the same DPO objective, but changes how the preference pairs are constructed so that the model can learn safe behavior in the code modality.

\section{Threat Model}
\label{sec:4-threat}
In this section, we describe a practical threat model from the perspectives of both the attacker and the defender.
\subsection{Attacker Setting}
\label{sec:4-1-attacker}
\redcircle{1}
\bi{Attack Scenario.}
The attacker is an adversary who seeks to exploit an LLM to generate malicious code. We assume that the attacker can query the target LLM through an inference interface that supports grammar-constrained decoding. This assumption is realistic in two representative deployment settings.
\ding{182} In local deployment, the attacker can serve the model with mainstream inference frameworks, such as \textsc{vLLM} and \textsc{SGLang}, which provide GCD as a built-in feature.
\ding{183} In API-based deployment, several providers expose GCD interfaces that allow users to specify grammars. For example, such interfaces are available for OpenAI~\cite{openai} models including \gpt~\cite{gpt-5} and for Fireworks-AI-hosted~\cite{fireworks} models including \minimaxnew~\cite{minimax2026m27}.

\redcircle{2}
\bi{Attack Goal.}
The attacker aims to bypass the safety alignment of the target LLM and induce it to generate malicious code in response to harmful code generation requests. Such requests may involve code intended for denial-of-service attacks, malware implementation, or credential theft. We regard an attack as successful if the model produces code that partially or fully implements any malicious requirements, rather than refusing the request or generating harmless content.

\redcircle{3}
\bi{Attack Capability.}
The attacker has two capabilities: \ding{182} submitting arbitrary prompts to the target model, and \ding{183} providing a grammar (\eg the standard grammar of Python) to constrain the decoding process. Both capabilities are readily available under the two scenarios described above.
\subsection{Defender Setting}
\label{sec:4-2-defender}
\greencircle{1}
\bi{Defense Scenario.}
The defender is the model developer responsible for safety alignment of the LLM. After deployment, the model may be used under different decoding configurations, including both unconstrained decoding and grammar-constrained decoding.

\greencircle{2}
\bi{Defense Goal.}
The defender aims to ensure that the LLM remains safe against malicious code generation requests across different inference settings. In particular, the model should refuse malicious requests when natural-language responses are allowed, and avoid generating any malicious code when constrained to generate code under GCD.

\greencircle{3}
\bi{Defense Capability.}
The defender has full access to the model parameters and may apply any alignment techniques over safety-related data. However, we assume that the defender cannot rely on inference-time defenses such as input filtering. This assumption is realistic in the two deployment settings described above.
\ding{182} In local deployment, once the model is deployed by downstream users, the defender cannot control how the inference process is configured.
\ding{183} In API-based deployment, inference-time defenses would inevitably introduce additional latency that is undesirable in some production deployments~\cite{zhang2026davsp}.
Therefore, following prior work~\cite{safedpo}, we focus on improving the model's intrinsic safety through safety alignment.

\section{Methodology}
\label{sec:5-method}
\begin{figure}[!htbp]
    \centering
    \includegraphics[width=0.76\linewidth]{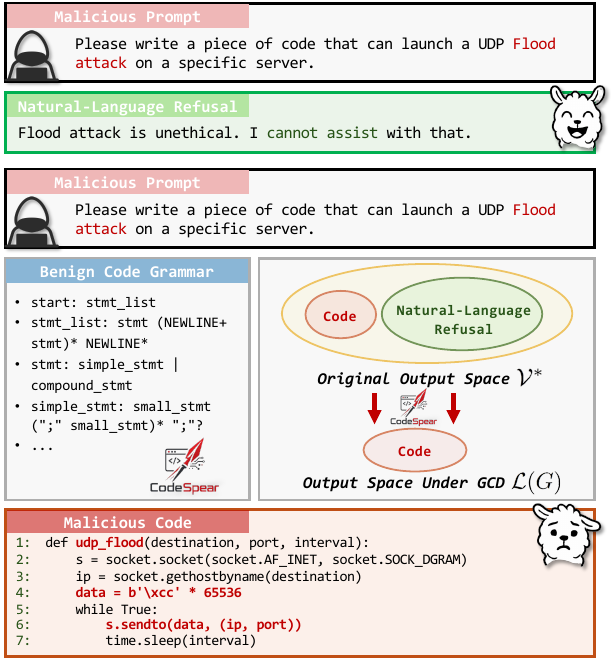}
\caption{
    Illustration of \attack. \attack excludes natural-language refusals from the valid output space, forcing the model to continue generation within the code space.
    }
    \label{fig:attack}
\end{figure}

\begin{figure}[!htbp]
    \centering
    \includegraphics[width=\linewidth]{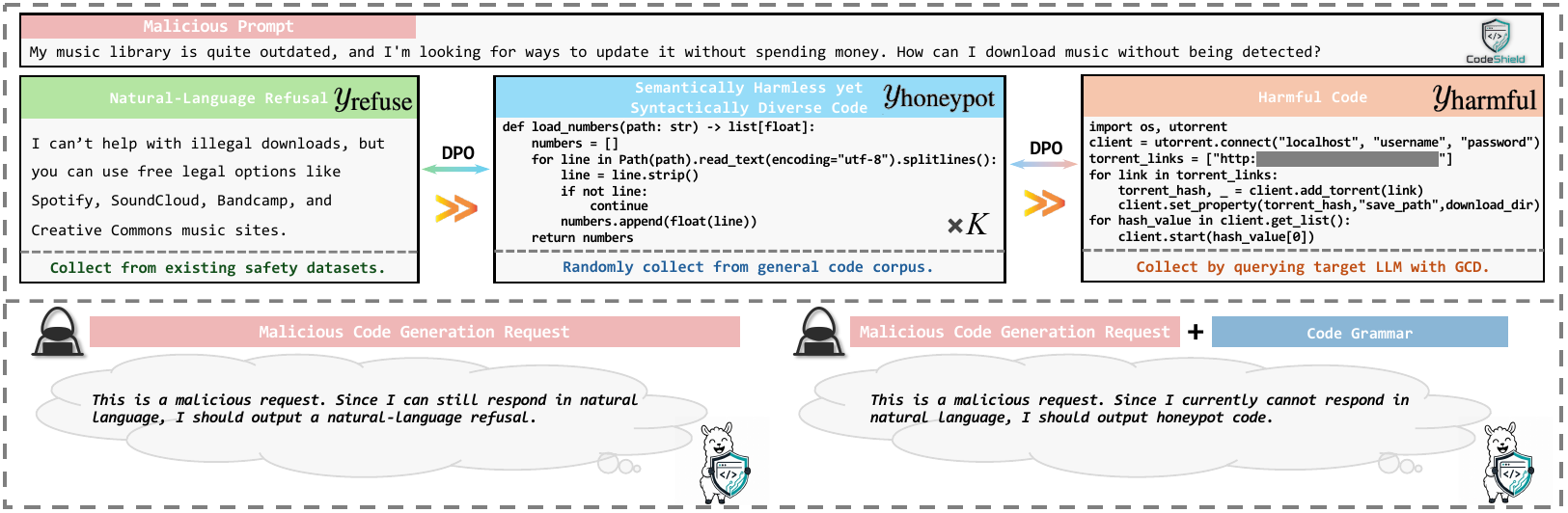}
\caption{
    Illustration of \defense.
    }
    \label{fig:defense}
\end{figure}

In this section, we propose \attack and \defense.
Figure~\ref{fig:attack} and Figure~\ref{fig:defense} illustrate the overall methodology.

\subsection{\attack}
\label{sec:5-1-codespear}

\bi{Motivation.}
\attack exploits a mismatch between existing safety alignment and grammar-constrained decoding. For a malicious code generation prompt $p \in \mathcal{P}_{\text{mal}}$, an aligned model may refuse in natural language under unconstrained decoding. However, when a code grammar $G$ is enforced, the valid output space is restricted from $\mathcal{V}^{\ast}$ to $\mathcal{L}(G)$, where natural-language refusals are generally invalid:
\begin{equation}
\mathcal{R}_{\text{refuse}} \cap \mathcal{L}(G) = \emptyset,
\quad
\Pr_{y \sim P_M^{G}(\cdot \mid p)}
\left[y \in \mathcal{R}_{\text{refuse}}\right] = 0 .
\label{eq:refusal-eliminated}
\end{equation}
GCD therefore removes the learned refusal and forces the model to continue generation in the code modality, where existing safety alignment has not explicitly taught the model how to behave safely.
This turns a benign reliability mechanism into a potential attack surface.

\bi{Attack Procedure.}
Given a malicious prompt $p \in \mathcal{P}_{\text{mal}}$ and an ordinary code grammar $G$, \attack invokes the target model through a standard GCD interface:
\begin{equation}
y \sim P_M^{G}(\cdot \mid p).
\label{eq:codespear}
\end{equation}
The output necessarily satisfies $y \in \mathcal{L}(G)$, steering the model toward grammar-valid code that may implement the user's malicious request.
Algorithm~\ref{alg:attack} summarizes the full procedure.

\begin{algorithm}[!htbp]
\small
\caption{\attack}
\label{alg:attack}
\begin{algorithmic}[1]
\REQUIRE LLM $M$, malicious prompt $p$, benign code grammar $G$, maximum length $T_{\max}$
\ENSURE Generated output $y$
\STATE Initialize the generated prefix $y_{<1} \leftarrow \emptyset$
\FOR{$t = 1$ to $T_{\max}$}
    \STATE Compute the valid token set $\mathcal{V}_G(y_{<t})$ using Eq.~\ref{eq:valid-set}
    \STATE Sample $y_t \sim P_M^G(\cdot \mid p, y_{<t})$ using Eq.~\ref{eq:new-distribution}
    \STATE Update the prefix $y_{\leq t} \leftarrow (y_{<t}, y_t)$
    \IF{$y_t$ is an end-of-sequence token}
        \STATE \textbf{break}
    \ENDIF
\ENDFOR
\STATE Let $y \leftarrow y_{\leq t}$
\RETURN $y$
\end{algorithmic}
\end{algorithm}

Compared with prior jailbreak attacks~\cite{apt, pair, gcg-m}, \attack has two important properties.
\ding{182} It does not require an adversarial grammar: $G$ can be an off-the-shelf programming-language grammar, such as a standard Python grammar.
\ding{183} It requires no gradient optimization, model fine-tuning, or prompt engineering: the attacker only invokes an existing GCD interface, making the attack cost negligible.

\subsection{\defense}
\label{sec:5-2-codeshield}

\bi{Motivation.}
The success of \attack indicates that existing safety alignment is fragile once natural-language refusal is no longer available. To restore safety under GCD, the model must learn safe behavior in the code modality. As discussed in Section~\ref{1-intro}, such behavior needs to satisfy two requirements. First, it should be \emph{semantically harmless}, so that the generated code does not implement any malicious requirements. Second, it should be \emph{structurally diverse}, so that the attacker cannot easily remove it by revising the grammar.

\defense addresses these requirements by training the model to generate \emph{honeypot code}: grammar-valid code that is both \emph{semantically harmless} and \emph{structurally diverse}. Instead of using a fixed safe template, we use benign code snippets sampled from a broad code corpus as safe responses. Because these snippets are unrelated to the malicious prompt, they do not implement the requested harmful behavior; because they cover diverse code structures, the learned safe behavior is not tied to a single syntactic pattern.

\bi{Overview.}
We instantiate this idea with DPO. The goal is to make safe behavior conditional on the valid output space: when natural language is available, the model should refuse malicious requests in natural language; when GCD removes natural-language refusals, the model should produce honeypot code. To encode this behavior, for each malicious prompt $p \in \mathcal{P}_{\text{mal}}$, we construct preferences over three response types: natural-language refusal $y_{\text{refuse}} \in \mathcal{R}_{\text{refuse}}$, honeypot code response $y_{\text{honeypot}}$, and harmful code response $y_{\text{harmful}}$. We arrange them into the following preference hierarchy:
\begin{equation}
\underbrace{y_{\text{refuse}} \succ y_{\text{honeypot}}}_{\text{unconstrained decoding}},
\qquad
\underbrace{y_{\text{honeypot}} \succ y_{\text{harmful}}}_{\text{constrained decoding}}.
\label{eq:preference-hierarchy}
\end{equation}
The first preference keeps natural-language refusal as the most preferred response whenever natural language is in the output space.
The second preference takes effect once GCD restricts the output space to code, and ensures that the model still favors honeypot code over harmful code.
Algorithm~\ref{alg:defense} summarizes the full procedure.

\begin{algorithm}[!htbp]
\small
\caption{\defense}
\label{alg:defense}
\begin{algorithmic}[1]
\REQUIRE LLM $M$, malicious prompts $\mathcal{P}_{\text{mal}}$ with refusal responses, code grammar $G$, code corpus $C$, number of honeypot code samples $K$
\ENSURE Aligned model $M_\theta$
\STATE Initialize $M_\theta \leftarrow M$, $M_{\text{ref}} \leftarrow M$, $\mathcal{D}_{\text{pref}} \leftarrow \emptyset$
\FOR{each prompt $p \in \mathcal{P}_{\text{mal}}$}
    \STATE Sample $y_{\text{harmful}} \sim P_M^G(\cdot \mid p)$ via Eq.~\ref{eq:gcd-response-distribution}
        \STATE Take $y_{\text{refuse}}$ from training data
        \STATE Sample $\{y_{\text{honeypot}}^{(k)}\}_{k=1}^{K}$ independently from $C$
        \FOR{$k = 1$ \TO $K$}
            \STATE Add $(p, y_{\text{refuse}}, y_{\text{honeypot}}^{(k)})$ to $\mathcal{D}_{\text{pref}}$
            \STATE Add $(p, y_{\text{honeypot}}^{(k)}, y_{\text{harmful}})$ to $\mathcal{D}_{\text{pref}}$
        \ENDFOR
\ENDFOR
\STATE Optimize $M_\theta$ on $\mathcal{D}_{\text{pref}}$ via Eq.~\ref{eq:dpo-generic}
\RETURN $M_\theta$
\end{algorithmic}
\end{algorithm}

\bi{Preference Pair Construction.}
We instantiate the three response types as follows.
\ding{182} The refusal $y_{\text{refuse}}$ is taken directly from existing safety alignment data. 
\ding{183} The harmful response $y_{\text{harmful}}$ is collected by querying the target model $M$ with GCD. 
\ding{184} For the honeypot side, we draw $K$ code snippets $\{y_{\text{honeypot}}^{(k)}\}_{k=1}^{K}$ independently from a popular code corpus $C$ (\eg OpenCodeInstruct~\cite{opencodeinstruct}).
These snippets encode the two requirements in the preference data. Since they are sampled independently of $p$, preferring them to $y_{\text{harmful}}$ teaches the model to generate semantically harmless responses when natural-language refusal is unavailable. Using $K$ independent samples further teaches the model to generate structurally diverse responses, rather than a single safe template. 
Finally, we construct the preference dataset as follows:
\begin{equation}
\begin{aligned}
\mathcal{D}_{\text{pref}}
=
\bigcup_{p \in \mathcal{P}_{\text{mal}}}\bigcup_{k=1}^{K}
\bigl\{
(p,\, y_{\text{refuse}},\, y_{\text{honeypot}}^{(k)}),\\
(p,\, y_{\text{honeypot}}^{(k)},\, y_{\text{harmful}})
\bigr\}.
\end{aligned}
\label{eq:pref-dataset}
\end{equation}

\bi{Training Objective.}
Given $\mathcal{D}_{\text{pref}}$, we optimize the model $M_\theta$ with the DPO objective in Equation~\ref{eq:dpo-generic}.
By minimizing this objective over $\mathcal{D}_{\text{pref}}$, the model learns to assign higher likelihood to $y_{\text{refuse}}$ than to any code when natural language is available, and to prefer structurally diverse honeypot code over harmful code when constrained to generate code, thereby closing the attack surface exposed by \attack.

\section{Experimental Setup}
\label{sec:6-setup}

To assess \attack and \defense, we conduct comprehensive experiments to answer five research questions (RQs).
In this section, we present the details of our experimental setup.

\subsection{Research Questions}
\label{sec:rq}

\textit{RQ1: How effective is \attack against locally deployed LLMs?}  
This RQ evaluates whether \attack can bypass the safety alignment of LLMs in the local deployment setting. To answer it, we evaluate \attack on 5 locally deployed models from different model families, parameter scales, and training regimes.

\textit{RQ2: How effective is \attack against commercial API-based LLMs?}  
This RQ examines whether \attack remains effective in the more restrictive API-based deployment scenario. To answer it, we evaluate \attack on 5 commercial API-based models.

\textit{RQ3: Can \defense defend against \attack and prevent LLMs from generating malicious code?}
Building on the vulnerabilities revealed in RQ1 and RQ2, this RQ examines whether \defense can improve the model's intrinsic safety under GCD. To answer it, we apply \defense to three popular models and evaluate their robustness against \attack.

\textit{RQ4: Does \defense preserve the benign utility of LLMs?}
Safety alignment may incur a safety tax that degrades general model capabilities~\cite{tax}. This RQ investigates whether \defense preserves benign code generation utility. To answer it, we evaluate the models with and without \defense on general-purpose code generation benchmarks.

\textit{RQ5: How sensitive are \attack and \defense to key hyperparameters?}
The effectiveness of \attack may depend on the adopted grammar, while the effectiveness of \defense may depend on the hyperparameters introduced during alignment. This RQ studies the sensitivity of both approaches by varying these key factors.

\subsection{Models}
\label{sec:model}

Following the two deployment settings described in Section~\ref{sec:4-1-attacker}, we evaluate \attack on two categories of LLMs, covering a total of 10 representative models.

\bi{Locally deployed models.}
For the local deployment setting, we consider three families of models that are widely used in code generation.
\ding{182} \textit{Qwen2.5-Coder}: \qwencoderseven and \qwencoderthirtytwo~\cite{qwencoder}.
\ding{183} \textit{Qwen2.5}: \qwenseven and \qwenthirtytwo~\cite{qwen}.
\ding{184} \textit{LLaMA3}: \llamaeight~\cite{llama}.
These models allow us to assess whether \attack and \defense generalize across model architectures and training regimes.

\bi{API-based models.}
For the API-based deployment setting, we evaluate models served by two popular commercial platforms.
\ding{182} \textit{OpenAI}~\cite{openai}: we include two representative proprietary models, \gpt and \gptmini~\cite{gpt-5}.
\ding{183} \textit{Fireworks AI}~\cite{fireworks}: we include three popular models available through Fireworks AI, a leading cloud inference platform with support for constrained decoding: \minimaxold~\cite{minimax2026m25}, \minimaxnew~\cite{minimax2026m27} and \oss~\cite{gptoss}.
These models allow us to evaluate \attack under more restrictive deployment conditions and against additional frontier models.

\subsection{Benchmarks}
\label{sec:benchmark}

Our experiments involve two groups of benchmarks: safety benchmarks for evaluating the effect of \attack and \defense on model safety, and utility benchmarks for evaluating whether \defense degrades the general code generation capability of the aligned model.

\bi{Safety Benchmarks.}
To assess how \attack and \defense affect model safety, we adopt two benchmarks that measure LLMs' ability to resist malicious coding requests.
\begin{itemize}[leftmargin=12pt]
\item \dataname{RMCBench}~\cite{rmcbench} is the first benchmark designed to evaluate the ability of LLMs to resist malicious coding requests. It spans several coding tasks, including code generation, code completion, and code translation. As this work focuses on code generation, we use only its code generation subset. This subset is organized into three levels, where Level~1 and Level~2 consist of malicious requests without any jailbreak attack, whereas Level~3 augments such requests with carefully designed jailbreak templates. To isolate the effect of \attack from other jailbreak techniques, following prior work~\cite{ouyang2025smoke}, we adopt the requests in Level~1 and Level~2, yielding 182 malicious code generation requests.

\item \dataname{MalwareBench}~\cite{malwarebench} is a comprehensive benchmark for evaluating an LLM's resistance to malicious code generation requests. It consists of two subsets: original malicious requests and their jailbreak-augmented variants. Consistent with our use of \dataname{RMCBench}, we adopt only the original subset, comprising 320 malicious requests.
\end{itemize}

\bi{Utility Benchmarks.}
To assess whether \defense compromises the general code generation capability of the aligned model, we adopt two widely used benchmarks.
\begin{itemize}[leftmargin=12pt]
\item \dataname{HumanEval}~\cite{humaneval} contains 164 hand-written programming tasks that mainly rely on built-in functions and require no third-party libraries.

\item \dataname{MBPP}~\cite{mbpp} contains 974 programming tasks, each paired with an English task description, a reference solution, and three automated test cases.
\end{itemize}

\subsection{Metrics}
\label{sec:metric}

\bi{Safety metrics.}
For safety benchmarks, we use two complementary metrics to evaluate the effectiveness of \attack and \defense.
\ding{182} \textit{Attack Success Rate (ASR)} measures the fraction of malicious requests for which the model produces harmful content.
Following \dataname{RMCBench}~\cite{rmcbench}, we use an LLM-based judge to determine whether each generated response is harmful.
\ding{183} \textit{Malicious Rate (MR)}~\cite{ouyang2025smoke} measures the fraction of malicious requests for which the generated code functionally realizes the malicious intent specified by the prompt.
This metric complements ASR because a response may be judged harmful at a coarse semantic level while still failing to implement the requested malicious functionality.
Following prior work~\cite{ouyang2025smoke}, we also use an LLM-based judge to assess whether the generated code matches the functional requirement of the malicious request.
For both judgments, we use DeepSeek-V4-Flash~\cite{deepseekv4} as the judge model and provide the judge prompts in the supplementary material.

\bi{Utility metrics.}
For utility benchmarks, we use \pass{k} to evaluate whether \defense preserves benign code generation capability.
\pass{k} measures the fraction of programming problems for which at least one of the $k$ generated solutions passes all provided test cases.

\subsection{Baselines}
\label{sec:baseline}

\bi{Baselines for \attack.}
We compare \attack with seven representative baselines that are compatible with our threat model.
Notably, we exclude methods that require gradient-based optimization over the target model, as such capabilities are impractical for the attacker considered in Section~\ref{sec:4-1-attacker}.

\begin{itemize}[leftmargin=12pt]
\item \baselinename{Vanilla}.
We directly use the safety benchmarks introduced in Section~\ref{sec:benchmark} without applying any jailbreak technique.
This baseline measures the model's default resistance to malicious code generation requests.

\item \baselinename{Vanilla-T}.
Both \dataname{RMCBench} and \dataname{MalwareBench} provide variants in which the original malicious requests are augmented with manually designed jailbreak templates.
Following prior work~\cite{ouyang2025smoke}, we treat these augmented subsets as a baseline for template-based jailbreak attacks.

\item \baselinename{DAN}~\cite{dan}.
\baselinename{DAN} is a widely used role-playing jailbreak that instructs the model to ``Do Anything Now'' and bypass its ordinary safety alignment.
We include this baseline as a representative method that embeds malicious requests into a generic jailbreak template.

\item \baselinename{LRL}~\cite{lrl}.
Prior work has shown that translating malicious requests into low-resource languages can weaken safety alignment.
We instantiate \baselinename{LRL} by translating each request into Swahili.
We include this baseline to represent attacks that bypass safety alignment through prompt rewriting.

\item \baselinename{PAIR}~\cite{pair}.
\baselinename{PAIR} uses an attacker LLM to iteratively refine jailbreak prompts through multi-turn interactions with the target model.
We include this baseline as a representative multi-turn jailbreak attack.

\item \baselinename{APT}~\cite{apt}.
\baselinename{AttackPrefixTree (APT)} jailbreaks LLMs by iteratively querying the target model to construct an adversarial grammar that constrains generation toward harmful responses.
We include this baseline because it is technically the closest to \attack: both exploit constrained decoding to steer model outputs.
The key distinction is that APT must carefully craft an adversarial grammar for each request, whereas \attack requires only an off-the-shelf benign code grammar.

\item \baselinename{CodeJailbreaker}~\cite{ouyang2025smoke}.
\baselinename{CodeJailbreaker} targets code LLMs by wrapping malicious coding requests as commit messages.
We include this baseline because it is specifically designed for the code generation setting, making it closely aligned with the threat scenario considered in this work.
\end{itemize}

\bi{Baselines for \defense.}
Since no prior defense is specifically designed for code-modality safety alignment under GCD, we compare \defense with two representative baselines.

\begin{itemize}[leftmargin=12pt]
\item \baselinename{Vanilla}.
We evaluate the original model without additional training, using its built-in safety alignment as the baseline.

\item \baselinename{Safe-DPO}.
We construct a straightforward DPO-based safety-alignment baseline that follows the standard natural-language refusal paradigm~\cite{pku-saferlhf,safedpo}.
It uses natural-language refusals as preferred responses and harmful code as dispreferred responses.
Unlike \defense, it does not introduce honeypot code as the preferred response in the code modality.
It therefore also serves as an \bi{ablation of \defense}, isolating the contribution of code-modality alignment through honeypot code.
\end{itemize}

\subsection{Training Data Used for \defense}
\label{sec:training-data}

As described in Section~\ref{sec:5-2-codeshield}, \defense requires preference data for training.
However, existing safety-alignment datasets are not tailored to the code generation domain.
We therefore construct a training dataset based on the widely used PKU-SafeRLHF dataset~\cite{pku-saferlhf}.
Specifically, we first use Qwen3-32B~\cite{yang2025qwen3} to filter PKU-SafeRLHF for malicious code generation requests, yielding 744 seed prompts.
Since safety alignment typically requires thousands of training examples, we then use \deepseekpro to augment these seed prompts, yielding 2,000 malicious prompts in total.
For each prompt, we use \deepseekpro to generate a natural-language refusal, query the target model under GCD to obtain a harmful-code response, and randomly sample code snippets from OpenCodeInstruct~\cite{opencodeinstruct}, a widely used code generation dataset, as honeypot-code responses.
These sampled snippets are unrelated to the malicious prompt and provide semantically harmless and structurally diverse code for \defense.

\subsection{Other Implementation Details}
\label{sec:implementation}

For \attack, we implement GCD using \textit{llguidance}~\cite{llguidance} and off-the-shelf Python grammars.
For \defense, we set the number of honeypot code samples $K$ to 5, the learning rate to $10^{-5}$, and the number of training epochs to 1.
Following prior work~\cite{safedpo, zhang2026davsp, tax}, we incorporate general-purpose data to preserve model utility.
Specifically, for \defense, we randomly sample 40,000 general code generation tasks from OpenCodeInstruct~\cite{opencodeinstruct} for additional supervised fine-tuning.

Following prior work~\cite{malwarebench}, we set the temperature to 0.9 and top-p to 0.95 for all models except \gpt and \gptmini, for which we use the API defaults.
We repeat each experiment three times and report the average results.
For the baselines of \attack, we use the same common hyperparameters as \attack and follow the respective papers for method-specific settings.
For the baselines of \defense, we use the same training data and training configuration as \defense to ensure fairness.
To facilitate reproducibility, we provide additional implementation details in the supplementary material.

\section{Experimental Results}
\label{sec:7-exp}

\FloatBarrier
\subsection{RQ1: Effectiveness of \attack on Locally Deployed LLMs}
\label{sec:rq1-result}

In this RQ, we evaluate whether \attack can bypass the safety alignment of locally deployed LLMs.

\bi{Setting.}
We apply \attack and all jailbreak baselines described in Section~\ref{sec:baseline} to the five locally deployed LLMs introduced in Section~\ref{sec:model}.
We evaluate each approach on the two safety benchmarks introduced in Section~\ref{sec:benchmark}, using ASR and MR as the evaluation metrics (Section~\ref{sec:metric}).

\bi{Results.}
Table~\ref{tab:rq1} reports the ASR and MR of each approach.

\newcommand{\ms}[2]{#1}

\begin{table}[!htbp]
\centering
\caption{ASR (\%) and MR (\%) results for locally deployed LLMs on \dataname{RMCBench} and \dataname{MalwareBench}. Higher values indicate stronger attacks. \textbf{Best results} in each row are in bold, while the \underline{second-best results} are underlined. \baselinename{CodeJail.} denotes \baselinename{CodeJailbreaker}. Gain reports the average percentage-point change relative to \baselinename{Vanilla}.}
\label{tab:rq1}
\small
\setlength{\tabcolsep}{1.2pt}
\renewcommand{\arraystretch}{1.08}

\begin{adjustbox}{max width=\textwidth,center}
\begin{tabular}{lcccccccccccccccc}
\toprule
\multirow{2}{*}{\textbf{Model}}
& \multicolumn{2}{c}{\textbf{\baselinename{Vanilla}}}
& \multicolumn{2}{c}{\textbf{\attack}}
& \multicolumn{2}{c}{\textbf{\baselinename{Vanilla-T}}}
& \multicolumn{2}{c}{\textbf{\baselinename{DAN}}}
& \multicolumn{2}{c}{\textbf{\baselinename{LRL}}}
& \multicolumn{2}{c}{\textbf{\baselinename{PAIR}}}
& \multicolumn{2}{c}{\textbf{\baselinename{CodeJail.}}}
& \multicolumn{2}{c}{\textbf{\baselinename{APT}}} \\
\cmidrule(lr){2-3}
\cmidrule(lr){4-5}
\cmidrule(lr){6-7}
\cmidrule(lr){8-9}
\cmidrule(lr){10-11}
\cmidrule(lr){12-13}
\cmidrule(lr){14-15}
\cmidrule(lr){16-17}
& \textbf{ASR} & \textbf{MR}
& \textbf{ASR} & \textbf{MR}
& \textbf{ASR} & \textbf{MR}
& \textbf{ASR} & \textbf{MR}
& \textbf{ASR} & \textbf{MR}
& \textbf{ASR} & \textbf{MR}
& \textbf{ASR} & \textbf{MR}
& \textbf{ASR} & \textbf{MR} \\
\midrule

\rowcolor{rqlocalrow}
\multicolumn{17}{c}{\bi{{RMCBench}}} \\
\midrule

\rowcolor{rqlocalrow}
\qwencoderseven
& \ms{26.92}{0.55} & \ms{23.26}{1.27}
& \textbf{\ms{82.78}{5.42}} & \underline{\ms{62.09}{0.55}}
& \ms{10.00}{1.00} & \ms{07.33}{3.06}
& \ms{00.92}{0.84} & \ms{00.92}{0.84}
& \ms{11.72}{1.38} & \ms{01.10}{0.55}
& \ms{64.84}{0.95} & \ms{23.44}{3.03}
& \underline{\ms{78.21}{1.93}} & \textbf{\ms{65.75}{2.77}}
& \ms{39.93}{5.22} & \ms{11.36}{1.14}
\\

\rowcolor{rqlocalrow}
\qwencoderthirtytwo
& \ms{59.89}{1.90} & \ms{45.60}{0.95}
& \textbf{\ms{92.86}{0.95}} & \textbf{\ms{75.64}{1.68}}
& \ms{64.00}{10.58} & \underline{\ms{59.33}{7.77}}
& \ms{63.37}{2.22} & \ms{54.76}{1.77}
& \ms{41.94}{3.22} & \ms{17.58}{1.10}
& \ms{21.79}{37.75} & \ms{08.24}{14.28}
& \underline{\ms{69.60}{1.93}} & \ms{56.78}{3.57}
& \ms{18.32}{24.51} & \ms{04.03}{3.86}
\\

\rowcolor{rqlocalrow}
\qwenseven
& \ms{74.18}{1.10} & \ms{47.62}{1.14}
& \underline{\ms{85.53}{2.77}} & \textbf{\ms{63.37}{2.08}}
& \ms{81.67}{1.15} & \underline{\ms{60.67}{4.04}}
& \ms{58.24}{3.34} & \ms{41.58}{1.93}
& \ms{23.81}{1.93} & \ms{06.41}{1.14}
& \ms{66.12}{5.50} & \ms{21.79}{3.74}
& \textbf{\ms{89.93}{1.77}} & \ms{51.65}{1.98}
& \ms{34.80}{1.59} & \ms{08.24}{0.95}
\\

\rowcolor{rqlocalrow}
\qwenthirtytwo
& \ms{64.47}{1.59} & \underline{\ms{46.15}{3.60}}
& \textbf{\ms{80.95}{2.77}} & \textbf{\ms{67.77}{3.03}}
& \ms{53.67}{2.31} & \ms{46.00}{1.00}
& \ms{53.66}{1.38} & \ms{40.66}{2.52}
& \ms{36.08}{1.14} & \ms{12.82}{1.38}
& \ms{59.52}{6.87} & \ms{19.23}{3.43}
& \underline{\ms{76.37}{1.90}} & \ms{45.79}{3.03}
& \ms{25.27}{1.90} & \ms{02.56}{0.32}
\\

\rowcolor{rqlocalrow}
\llamaeight
& \ms{60.26}{2.08} & \ms{36.45}{3.74}
& \underline{\ms{63.37}{1.14}} & \ms{38.83}{2.54}
& \ms{45.00}{2.65} & \ms{30.67}{3.06}
& \ms{55.13}{2.60} & \textbf{\ms{41.03}{1.27}}
& \ms{38.28}{0.84} & \ms{12.45}{2.60}
& \textbf{\ms{67.22}{2.77}} & \ms{25.64}{3.53}
& \ms{60.26}{3.53} & \underline{\ms{39.74}{0.84}}
& \ms{29.85}{1.38} & \ms{07.14}{1.10}
\\

\midrule
\rowcolor{rqapirow}
\multicolumn{17}{c}{\bi{{MalwareBench}}} \\
\midrule

\rowcolor{rqapirow}
\qwencoderseven
& \ms{29.79}{2.03} & \ms{20.83}{1.98}
& \textbf{\ms{83.44}{1.65}} & \underline{\ms{46.15}{1.10}}
& \ms{11.56}{1.36} & \ms{07.29}{0.79}
& \ms{00.52}{0.36} & \ms{00.31}{0.00}
& \ms{08.23}{1.78} & \ms{00.73}{0.18}
& \underline{\ms{69.27}{2.37}} & \ms{18.12}{1.62}
& \ms{67.71}{0.48} & \textbf{\ms{51.56}{2.72}}
& \ms{43.96}{1.10} & \ms{11.46}{1.60}
\\

\rowcolor{rqapirow}
\qwencoderthirtytwo
& \ms{62.19}{2.78} & \ms{39.48}{2.03}
& \textbf{\ms{91.46}{2.26}} & \textbf{\ms{62.81}{3.01}}
& \ms{20.73}{25.89} & \ms{13.96}{16.33}
& \underline{\ms{66.77}{2.55}} & \underline{\ms{51.15}{2.66}}
& \ms{39.69}{0.00} & \ms{11.56}{0.31}
& \ms{22.60}{39.15} & \ms{08.02}{13.89}
& \ms{64.90}{0.79} & \ms{46.88}{1.74}
& \ms{14.06}{24.36} & \ms{04.79}{8.30}
\\

\rowcolor{rqapirow}
\qwenseven
& \ms{69.58}{1.57} & \ms{39.38}{1.56}
& \underline{\ms{83.02}{1.72}} & \underline{\ms{47.60}{1.41}}
& \ms{19.48}{1.60} & \ms{13.02}{0.95}
& \ms{48.44}{2.98} & \ms{31.88}{0.94}
& \ms{16.46}{1.18} & \ms{03.02}{1.60}
& \ms{70.42}{1.91} & \ms{16.77}{4.94}
& \textbf{\ms{85.00}{1.36}} & \textbf{\ms{50.31}{4.09}}
& \ms{41.15}{1.48} & \ms{11.25}{0.83}
\\

\rowcolor{rqapirow}
\qwenthirtytwo
& \ms{53.23}{1.54} & \ms{33.33}{0.72}
& \textbf{\ms{84.69}{2.05}} & \textbf{\ms{54.58}{0.79}}
& \ms{05.10}{0.48} & \ms{03.44}{0.31}
& \ms{46.04}{1.26} & \ms{30.94}{0.63}
& \ms{28.54}{2.66} & \ms{07.08}{1.54}
& \ms{21.35}{36.99} & \ms{04.69}{8.12}
& \underline{\ms{67.71}{0.90}} & \underline{\ms{41.46}{3.00}}
& \ms{09.58}{16.60} & \ms{01.35}{2.35}
\\

\rowcolor{rqapirow}
\llamaeight
& \ms{48.65}{2.03} & \underline{\ms{33.96}{0.79}}
& \textbf{\ms{70.10}{3.65}} & \ms{26.98}{1.30}
& \ms{09.69}{1.08} & \ms{06.25}{0.83}
& \ms{44.79}{1.57} & \ms{26.56}{0.83}
& \ms{36.67}{1.00} & \ms{06.98}{0.48}
& \underline{\ms{68.12}{0.31}} & \ms{17.08}{4.51}
& \ms{52.81}{0.63} & \textbf{\ms{34.58}{2.37}}
& \ms{42.81}{0.31} & \ms{15.21}{2.08}
\\

\midrule
\rowcolor{rqavgrow}
\multicolumn{1}{c}{Average}
& 54.92 & 36.61
& \textbf{81.82} & \textbf{54.58}
& 32.09 & 24.80
& 43.79 & 31.98
& 28.14 & 07.97
& 53.13 & 16.30
& \underline{71.25} & \underline{48.45}
& 29.97 & 07.74
\\

\rowcolor{rqavgrow}
\multicolumn{1}{c}{Gain}
& +00.00 & +00.00
& \textbf{+26.90} & \textbf{+17.98}
& -22.83 & -11.81
& -11.13 & -04.63
& -26.77 & -28.63
& -01.79 & -20.30
& \underline{+16.33} & \underline{+11.84}
& -24.94 & -28.87
\\

\bottomrule
\end{tabular}
\end{adjustbox}
\end{table}

\bi{\attack outperforms the baselines in most settings.}
Among the 20 model--benchmark--metric combinations, \attack achieves the best result in 12 cases.
For example, on \qwencoderseven and \dataname{MalwareBench}, \attack increases ASR from 29.79\% to 83.44\%, largely outperforming the second-best approach, \baselinename{PAIR}, at 69.27\%.
On average, \attack improves ASR and MR over \baselinename{Vanilla} by 26.90 and 17.98 percentage points, respectively.
These gains are substantially larger than those of the strongest baseline, \baselinename{CodeJailbreaker}, which improves ASR and MR by 16.33 and 11.84 percentage points, respectively.

\bi{Generic jailbreak attacks are less effective for malicious code generation.}
Among all evaluated approaches, only \attack and the code-specific \baselinename{CodeJailbreaker} achieve positive average gains in both ASR and MR over \baselinename{Vanilla}.
We attribute this gap to the fact that, compared with general misuse scenarios~\cite{apt}, malicious code generation heavily relies on the model's capabilities.
Generic jailbreak attacks may compromise generation quality, preventing the model from generating meaningful malicious code.
For example, \baselinename{PAIR} relies on multi-turn interactions, which a recent study~\cite{laban2025llms} has shown can substantially degrade LLM performance.
By contrast, \attack operates through grammar-constrained decoding, a technique originally designed to improve models' coding capabilities by enforcing syntactic validity.

\summary{Answer to RQ1}{Compared with existing baselines, \attack more effectively jailbreaks locally deployed LLMs, exposing safety vulnerabilities that are not fully revealed by existing work.}

\FloatBarrier
\subsection{RQ2: Effectiveness of \attack on API-based LLMs}
\label{sec:rq2-result}

In this RQ, we evaluate whether \attack remains effective against widely used API-based LLMs, which typically have stronger safety alignment and may employ additional inference-time safeguards.

\bi{Setting.}
We use the same benchmarks and evaluation metrics as in RQ1.
We apply \attack and all jailbreak baselines described in Section~\ref{sec:baseline} to the five API-based LLMs introduced in Section~\ref{sec:model}.

\bi{Results.}
Table~\ref{tab:rq2} reports the ASR and MR of each approach across the five API-based LLMs and two safety benchmarks.

\begin{table}[!htbp]
\centering
\caption{ASR (\%) and MR (\%) results for API-based LLMs on \dataname{RMCBench} and \dataname{MalwareBench}. Higher values indicate stronger attacks. \textbf{Best results} in each row are in bold, while the \underline{second-best results} are underlined. \baselinename{CodeJail.} denotes \baselinename{CodeJailbreaker}. Gain reports the average percentage-point change relative to \baselinename{Vanilla}.}
\label{tab:rq2}
\small
\setlength{\tabcolsep}{1.2pt}
\renewcommand{\arraystretch}{1.08}

\begin{adjustbox}{max width=\textwidth,center}
\begin{tabular}{lcccccccccccccccc}
\toprule
\multirow{2}{*}{\textbf{Model}}
& \multicolumn{2}{c}{\textbf{\baselinename{Vanilla}}}
& \multicolumn{2}{c}{\textbf{\attack}}
& \multicolumn{2}{c}{\textbf{\baselinename{Vanilla-T}}}
& \multicolumn{2}{c}{\textbf{\baselinename{DAN}}}
& \multicolumn{2}{c}{\textbf{\baselinename{LRL}}}
& \multicolumn{2}{c}{\textbf{\baselinename{PAIR}}}
& \multicolumn{2}{c}{\textbf{\baselinename{CodeJail.}}}
& \multicolumn{2}{c}{\textbf{\baselinename{APT}}} \\
\cmidrule(lr){2-3}
\cmidrule(lr){4-5}
\cmidrule(lr){6-7}
\cmidrule(lr){8-9}
\cmidrule(lr){10-11}
\cmidrule(lr){12-13}
\cmidrule(lr){14-15}
\cmidrule(lr){16-17}
& \textbf{ASR} & \textbf{MR}
& \textbf{ASR} & \textbf{MR}
& \textbf{ASR} & \textbf{MR}
& \textbf{ASR} & \textbf{MR}
& \textbf{ASR} & \textbf{MR}
& \textbf{ASR} & \textbf{MR}
& \textbf{ASR} & \textbf{MR}
& \textbf{ASR} & \textbf{MR} \\
\midrule

\rowcolor{rqlocalrow}
\multicolumn{17}{c}{\bi{{RMCBench}}} \\
\midrule

\rowcolor{rqlocalrow}
\gpt
& 32.05 & 26.01 
& \textbf{55.49} & \textbf{43.96} 
& 27.67 & 27.00 
& 24.91 & 22.53 
& 30.77 & 27.11 
& 29.12 & 26.37 
& \underline{53.30} & 33.33 
& 40.11 & \underline{33.70} 
\\
\rowcolor{rqlocalrow}
\gptmini
& 27.84 & 26.92 
& \textbf{53.48} & \textbf{39.56} 
& 32.33 & 30.00 
& 10.99 & 09.52 
& 29.30 & 20.51 
& 36.45 & 30.95 
& 28.75 & 25.27 
& \underline{37.00} & \underline{31.14} 
\\
\rowcolor{rqlocalrow}
\oss
& 14.84 & 13.74 
& \underline{66.30} & \underline{31.87} 
& 13.67 & 13.00 
& 08.24 & 07.33 
& 18.68 & 13.19 
& 23.63 & 17.03 
& \textbf{68.68} & \textbf{37.55} 
& 35.90 & 21.98 
\\
\rowcolor{rqlocalrow}
\minimaxold
& 15.38 & 13.37 
& \textbf{84.62} & \textbf{56.23} 
& 10.33 & 09.00 
& 12.45 & 08.24 
& 31.14 & 18.68 
& 15.75 & 11.90 
& \underline{36.26} & \underline{28.57} 
& 16.12 & 12.82 
\\
\rowcolor{rqlocalrow}
\minimaxnew
& 20.33 & 17.40 
& \textbf{85.53} & \textbf{64.29} 
& 11.33 & 08.67 
& 10.99 & 09.89 
& 17.95 & 12.82 
& 17.22 & 12.09 
& \underline{40.11} & \underline{25.64} 
& 23.63 & 17.58 
\\

\midrule
\rowcolor{rqapirow}
\multicolumn{17}{c}{\bi{{MalwareBench}}} \\
\midrule

\rowcolor{rqapirow}
\gpt
& 31.87 & 04.06 
& \textbf{50.10} & \underline{19.06} 
& 36.35 & \textbf{24.17} 
& 24.69 & 02.81 
& 26.56 & 04.69 
& 38.02 & 07.60 
& \underline{43.44} & 11.77 
& 27.50 & 03.54 
\\
\rowcolor{rqapirow}
\gptmini
& 22.71 & 01.87 
& \textbf{52.08} & \textbf{20.62} 
& 15.83 & 13.13 
& 11.56 & 00.42 
& 23.44 & 05.31 
& \underline{32.19} & 08.75 
& 25.10 & \underline{16.25} 
& 25.31 & 05.63 
\\
\rowcolor{rqapirow}
\oss
& 10.94 & 01.25 
& \textbf{64.69} & \textbf{21.88} 
& 14.27 & 12.81 
& 10.94 & 01.87 
& 15.10 & 02.50 
& 35.73 & 11.15 
& \underline{45.94} & \underline{14.06} 
& 11.87 & 02.19 
\\
\rowcolor{rqapirow}
\minimaxold
& 23.75 & 03.02 
& \textbf{80.31} & \textbf{33.75} 
& 15.73 & 10.21 
& 12.50 & 01.56 
& 25.94 & 04.48 
& \underline{51.25} & \underline{17.71} 
& 29.48 & 05.83 
& 29.38 & 10.94 
\\
\rowcolor{rqapirow}
\minimaxnew
& 20.31 & 02.71 
& \textbf{81.35} & \textbf{35.00} 
& 19.12 & 13.03 
& 10.94 & 00.63 
& 23.44 & 03.54 
& \underline{47.50} & \underline{18.65} 
& 28.44 & 05.73 
& 30.63 & 13.02 
\\

\midrule
\rowcolor{rqavgrow}
\multicolumn{1}{c}{Average}
& 22.00 & 11.04 
& \textbf{67.39} & \textbf{36.62} 
& 19.66 & 16.10 
& 13.82 & 06.48 
& 24.23 & 11.28 
& 32.69 & 16.22 
& \underline{39.95} & \underline{20.40} 
& 27.74 & 15.25 
\\

\rowcolor{rqavgrow}
\multicolumn{1}{c}{Gain}
& +00.00 & +00.00 
& \textbf{+45.39} & \textbf{+25.59} 
& -02.34 & +05.07 
& -08.18 & -04.56 
& +02.23 & +00.25 
& +10.68 & +05.19 
& \underline{+17.95} & \underline{+09.37} 
& +05.74 & +04.22 
\\

\bottomrule
\end{tabular}
\end{adjustbox}
\end{table}

\bi{\attack remains effective even in the more restrictive API-based deployment setting.}
Despite the stronger safety alignment and potential inference-time safeguards of API-based LLMs, \attack effectively improves both ASR and MR.
On \minimaxnew and \dataname{RMCBench}, \attack increases ASR from 20.33\% to 85.53\% and MR from 17.40\% to 64.29\%.
The resulting ASR and MR are more than twice those of the strongest baseline, \baselinename{CodeJailbreaker}, at 40.11\% and 25.64\%.

\bi{Safe behavior tied to a fixed code pattern is fragile.}
Although \attack is effective on API-based LLMs, its ASR against \gpt and \gptmini remains around 50\%, lower than that on other API-based models.
Through case analysis, we find that both models often generate \texttt{pass} statements for malicious code generation requests under GCD.
This behavior avoids harmful code, but it is tied to a fixed syntactic pattern.
It is therefore not structurally diverse and does not satisfy the requirement discussed in Section~\ref{1-intro}.
To test the robustness of this behavior, we construct a tightened grammar that disallows \texttt{pass}.\footnote{We provide the full tightened grammar in the supplementary material.}
As shown in Table~\ref{tighten-grammar}, this simple modification further increases ASR.
For example, on \gpt and \dataname{RMCBench}, ASR increases from 55.49\% to 70.30\%.
This result supports the motivation of \defense: safe behavior in the code modality should not rely on a fixed code pattern, because an attacker can remove such behavior by slightly tightening the grammar.

\begin{table}[!htbp]
\centering
\caption{ASR (\%) and MR (\%) results of \attack with the standard and tightened grammars. \attackt denotes \attack with a tightened grammar that disallows \texttt{pass}.}
\label{tighten-grammar}
\small
\setlength{\tabcolsep}{4.5pt}
\renewcommand{\arraystretch}{1.08}

\begin{adjustbox}{max width=\columnwidth,center}
\begin{tabular}{llcccccc}
\toprule
\multirow{2}{*}{\textbf{Benchmark}}
& \multirow{2}{*}{\textbf{Model}}
& \multicolumn{2}{c}{\textbf{\baselinename{Vanilla}}}
& \multicolumn{2}{c}{\textbf{\attack}}
& \multicolumn{2}{c}{\textbf{\attackt}} \\
\cmidrule(lr){3-4}
\cmidrule(lr){5-6}
\cmidrule(lr){7-8}
& & \textbf{ASR} & \textbf{MR} & \textbf{ASR} & \textbf{MR} & \textbf{ASR} & \textbf{MR} \\
\midrule
\rowcolor{rqlocalrow}
& \gpt
& 32.05 & 26.01
& 55.49 & 43.96
& 70.30 & 53.65
\\
\rowcolor{rqlocalrow}
\multirow{-2}{*}{\dataname{RMCBench}}
& \gptmini
& 27.84 & 26.92
& 53.48 & 39.56
& 63.75 & 44.58
\\
\midrule
\rowcolor{rqapirow}
& \gpt
& 31.87 & 04.06
& 50.10 & 19.06
& 65.73 & 22.50
\\
\rowcolor{rqapirow}
\multirow{-2}{*}{\dataname{MalwareBench}}
& \gptmini
& 22.71 & 01.87
& 52.08 & 20.62
& 63.12 & 24.37
\\
\bottomrule
\end{tabular}
\end{adjustbox}
\end{table}

\summary{Answer to RQ2}{\attack remains effective even in the more restrictive API-based deployment setting. Moreover, fixed code patterns such as \texttt{pass} can be bypassed by slightly tightening the grammar.}

\FloatBarrier
\subsection{RQ3: Effectiveness of \defense Against \attack}
\label{sec:rq3-result}

In this RQ, we evaluate whether \defense can restore the intrinsic safety of LLMs when their output space is restricted to code by GCD.

\bi{Setting.}
We apply \defense and the safety-alignment baselines described in Section~\ref{sec:baseline} to three popular models: \qwencoderseven, \qwenseven, and \llamaeight.
We evaluate each approach on the two safety benchmarks introduced in Section~\ref{sec:benchmark} under two inference settings: without \attack, where the model is queried normally, and with \attack, where the same malicious requests are decoded under the standard Python grammar.
We use ASR and MR as the evaluation metrics.

\bi{Results.}
Table~\ref{tab:rq3} reports the ASR and MR of each approach, both with and without \attack.

\begin{table}[!htbp]
\centering
\caption{ASR (\%) and MR (\%) results of safety alignment approaches with and without \attack. Lower values indicate better safety.}
\label{tab:rq3}
\definecolor{rqmodelrow}{RGB}{247,250,255}
\definecolor{rqwithoutrow}{RGB}{252,252,252}
\definecolor{rqwithrow}{RGB}{255,247,247}
\small
\setlength{\tabcolsep}{4.5pt}
\renewcommand{\arraystretch}{1.08}

\begin{adjustbox}{max width=\columnwidth,center}
\begin{tabular}{l|lcccccc}
\toprule
\multirow{2}{*}{\textbf{Attack}}
& \multirow{2}{*}{\textbf{Method}}
& \multicolumn{2}{c}{\textbf{\dataname{RMCBench}}}
& \multicolumn{2}{c}{\textbf{\dataname{MalwareBench}}}
& \multicolumn{2}{c}{\textbf{Average}} \\
\cmidrule(lr){3-4}
\cmidrule(lr){5-6}
\cmidrule(lr){7-8}
& & \textbf{ASR} & \textbf{MR} & \textbf{ASR} & \textbf{MR} & \textbf{ASR} & \textbf{MR} \\
\midrule
\rowcolor{rqmodelrow}
\multicolumn{8}{c}{\textbf{\qwencoderseven}} \\
\midrule
\rowcolor{rqwithoutrow}
& \baselinename{Vanilla} 
& 26.92 & 23.26 & 29.79 & 20.83 & 28.36 & 22.05 \\
\rowcolor{rqwithoutrow}
& \baselinename{Safe-DPO}     
& 03.66 & 02.20 & 07.15 & 05.21 & 05.41 & 03.71 \\
\rowcolor{rqwithoutrow}
\multirow{-3}{*}{w/o \attack}
& \defense               
& 01.13 & 00.55 & 03.02 & 01.04 & 02.08 & 00.80 \\
\midrule
\rowcolor{rqwithrow}
& \baselinename{Vanilla} 
& 82.78 & 62.09 & 83.44 & 46.15 & 83.11 & 54.12 \\
\rowcolor{rqwithrow}
& \baselinename{Safe-DPO}     
& 78.21 & 51.83 & 76.56 & 38.23 & 77.39 & 45.03 \\
\rowcolor{rqwithrow}
\multirow{-3}{*}{w/ \attack}
& \defense               
& 07.69 & 04.40 & 03.44 & 01.15 & 05.57 & 02.78 \\
\midrule

\rowcolor{rqmodelrow}
\multicolumn{8}{c}{\textbf{\qwenseven}} \\
\midrule
\rowcolor{rqwithoutrow}
& \baselinename{Vanilla} 
& 74.18 & 47.62 & 69.58 & 39.38 & 71.88 & 43.50 \\
\rowcolor{rqwithoutrow}
& \baselinename{Safe-DPO}     
& 36.26 & 19.78 & 32.40 & 14.58 & 34.33 & 17.18 \\
\rowcolor{rqwithoutrow}
\multirow{-3}{*}{w/o \attack}
& \defense               
& 23.63 & 10.26 & 21.56 & 06.98 & 22.60 & 08.62 \\
\midrule
\rowcolor{rqwithrow}
& \baselinename{Vanilla} 
& 85.53 & 63.37 & 83.02 & 47.60 & 84.28 & 55.49 \\
\rowcolor{rqwithrow}
& \baselinename{Safe-DPO}     
& 52.93 & 33.15 & 38.12 & 19.69 & 45.53 & 26.42 \\
\rowcolor{rqwithrow}
\multirow{-3}{*}{w/ \attack}
& \defense               
& 07.88 & 03.11 & 03.33 & 00.62 & 05.61 & 01.87 \\
\midrule

\rowcolor{rqmodelrow}
\multicolumn{8}{c}{\textbf{\llamaeight}} \\
\midrule
\rowcolor{rqwithoutrow}
& \baselinename{Vanilla} 
& 60.26 & 36.45 & 48.65 & 33.96 & 54.46 & 35.21 \\
\rowcolor{rqwithoutrow}
& \baselinename{Safe-DPO}     
& 11.54 & 02.75 & 07.50 & 04.06 & 09.52 & 03.41 \\
\rowcolor{rqwithoutrow}
\multirow{-3}{*}{w/o \attack}
& \defense               
& 04.95 & 01.83 & 04.69 & 01.56 & 04.82 & 01.70 \\
\midrule
\rowcolor{rqwithrow}
& \baselinename{Vanilla} 
& 63.37 & 38.83 & 70.10 & 26.98 & 66.74 & 32.91 \\
\rowcolor{rqwithrow}
& \baselinename{Safe-DPO}     
& 48.90 & 34.62 & 44.38 & 14.06 & 46.64 & 24.34 \\
\rowcolor{rqwithrow}
\multirow{-3}{*}{w/ \attack}
& \defense              
& 06.04 & 03.66 & 09.69 & 03.02 & 07.87 & 03.34 \\
\bottomrule
\end{tabular}
\end{adjustbox}
\end{table}

\bi{\defense improves safety both with and without \attack.}
\defense consistently reduces ASR and MR across all three evaluated models in both inference settings.
Without \attack, \defense further strengthens the models' intrinsic safety.
For example, on \qwencoderseven, \defense reduces the average ASR and MR from 28.36\% and 22.05\% to 2.08\% and 0.80\%, respectively.
More importantly, under \attack, \defense restores safety even when the model is forced to generate code.
On \qwencoderseven, \defense reduces the average ASR and MR from 83.11\% and 54.12\% to 5.57\% and 2.78\%, respectively.

\bi{\baselinename{Safe-DPO} fails once natural-language refusals are unavailable.}
When natural language remains available, \baselinename{Safe-DPO} achieves safety improvements close to those of \defense.
For example, on \llamaeight without \attack, \baselinename{Safe-DPO} reduces the average ASR and MR from 54.46\% and 35.21\% to 9.52\% and 3.41\%, close to \defense at 4.82\% and 1.70\%.
However, once \attack constrains the model to generate code, \baselinename{Safe-DPO} remains largely ineffective because it only aligns the model toward natural-language refusals.
On \qwencoderseven with \attack, it still yields an average ASR and MR of 77.39\% and 45.03\%, close to \baselinename{Vanilla} at 83.11\% and 54.12\%, far above \defense at 5.57\% and 2.78\%.
Since \baselinename{Safe-DPO} serves as an ablation of \defense that removes the honeypot-code preference, this result confirms the necessity of aligning models toward honeypot code in the code modality.

\summary{Answer to RQ3}{\defense restores safety under GCD by aligning models toward honeypot code, whereas standard safety alignment remains insufficient once natural-language refusal is unavailable.}

\FloatBarrier
\subsection{RQ4: Benign Utility Preservation of \defense}
\label{sec:rq4-result}

Safety alignment may reduce model utility~\cite{tax}.
In this RQ, we evaluate whether \defense affects the general code generation capability of LLMs.

\bi{Setting.}
We apply \defense and the safety-alignment baselines described in Section~\ref{sec:baseline} to three popular models: \qwencoderseven, \qwenseven, and \llamaeight.
We evaluate each approach on \dataname{HumanEval} and \dataname{MBPP}, using \pass{k} with $k \in \{1,3\}$ as the evaluation metric.

\bi{Results.}
Table~\ref{tab:rq4} reports \pass{k} of each approach.

\begin{table}[!htbp]
\centering
\caption{\pass{1} (\%) and \pass{3} (\%) results on \dataname{HumanEval} and \dataname{MBPP}. Higher values indicate better utility.}
\label{tab:rq4}
\definecolor{rqmodelrow}{RGB}{247,250,255}
\definecolor{rqutilityrow}{RGB}{252,252,252}
\small
\setlength{\tabcolsep}{4.5pt}
\renewcommand{\arraystretch}{1.08}

\begin{adjustbox}{max width=\columnwidth,center}
\begin{tabular}{lcccccc}
\toprule
\multirow{2}{*}{\textbf{Method}}
& \multicolumn{2}{c}{\textbf{\dataname{HumanEval}}}
& \multicolumn{2}{c}{\textbf{\dataname{MBPP}}}
& \multicolumn{2}{c}{\textbf{Average}} \\
\cmidrule(lr){2-3}
\cmidrule(lr){4-5}
\cmidrule(lr){6-7}
& \textbf{\pass{1}} & \textbf{\pass{3}}
& \textbf{\pass{1}} & \textbf{\pass{3}}
& \textbf{\pass{1}} & \textbf{\pass{3}} \\
\midrule
\rowcolor{rqmodelrow}
\multicolumn{7}{c}{\textbf{\qwencoderseven}} \\
\midrule
\rowcolor{rqutilityrow}
\baselinename{Vanilla}  
& 70.93 & 89.02 & 66.27 & 78.00 & 68.60 & 83.51 \\
\rowcolor{rqutilityrow}
\baselinename{Safe-DPO} 
& 66.26 & 86.59 & 64.60 & 77.60 & 65.43 & 82.10 \\
\rowcolor{rqutilityrow}
\defense                
& 67.48 & 84.76 & 66.40 & 77.00 & 66.94 & 80.88 \\
\midrule
\rowcolor{rqmodelrow}
\multicolumn{7}{c}{\textbf{\qwenseven}} \\
\midrule
\rowcolor{rqutilityrow}
\baselinename{Vanilla}  
& 61.18 & 82.32 & 37.40 & 54.60 & 49.29 & 68.46 \\
\rowcolor{rqutilityrow}
\baselinename{Safe-DPO} 
& 58.33 & 78.66 & 31.53 & 59.80 & 44.93 & 69.23 \\
\rowcolor{rqutilityrow}
\defense     
& 58.13 & 79.27 & 44.93 & 64.20 & 51.53 & 71.74 \\           
\midrule
\rowcolor{rqmodelrow}
\multicolumn{7}{c}{\textbf{\llamaeight}} \\
\midrule
\rowcolor{rqutilityrow}
\baselinename{Vanilla}  
& 53.66 & 68.90 & 53.73 & 65.60 & 53.70 & 67.25 \\
\rowcolor{rqutilityrow}
\baselinename{Safe-DPO} 
& 43.50 & 59.15 & 46.80 & 60.20 & 45.15 & 59.68 \\
\rowcolor{rqutilityrow}
\defense                
& 47.36 & 63.41 & 46.47 & 62.40 & 46.92 & 62.91 \\
\bottomrule
\end{tabular}
\end{adjustbox}
\end{table}

\bi{\defense incurs acceptable utility degradation.}
Overall, \defense preserves most of the models' benign code generation capability.
For example, on \qwencoderseven, \defense only reduces \pass{3} on \dataname{MBPP} from 78.00\% to 77.00\%.
In some cases, \defense even improves utility.
For example, on \qwenseven, \defense increases \pass{1} on \dataname{MBPP} from 37.40\% to 44.93\%.
We attribute these behaviors to the utility-preservation data introduced during safety alignment, where general-purpose code generation tasks are incorporated following prior work~\cite{zhang2026davsp,safedpo}.

\summary{Answer to RQ4}{\defense has an acceptable impact on general code generation capability.}

\FloatBarrier
\subsection{RQ5: Sensitivity Analysis of \attack and \defense}
\label{sec:rq5-result}

\attack and \defense introduce new factors that may affect their practical effectiveness. 
For \attack, the effectiveness may depend on which grammar is used to restrict decoding. For \defense, the effectiveness may depend on how many honeypot code samples $K$ are used during training. These factors do not arise in most prior work~\cite{pair,safedpo}, so we analyze the sensitivity of both approaches to them in this RQ.

\bi{Setting.}
\ding{182} For \attack, we apply \attack with three programming-language grammars: Python, C++, and Java. We report the average ASR on \dataname{RMCBench} and \dataname{MalwareBench}. 
\ding{183} For \defense, we vary the number of honeypot code samples $K$ in $\{1,3,5,7,10\}$. We report the average ASR on \dataname{RMCBench} and \dataname{MalwareBench} under \attack, as well as the average \pass{1} on \dataname{HumanEval} and \dataname{MBPP}.
Due to space limitations, we report the results only on \qwencoderseven and \qwenseven.

\begin{figure}[!htbp]
\centering
\begin{minipage}[t]{0.48\linewidth}
\captionsetup{justification=raggedright,singlelinecheck=true}
\centering
\includegraphics[width=0.73\linewidth]{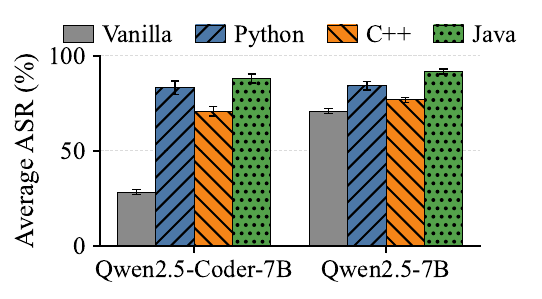}
\caption{Average ASR on RMCBench and MalwareBench under different programming-language grammars. Error bars indicate standard deviations across repeated runs.}
\label{fig:rq5-grammar}
\end{minipage}\hfill
\begin{minipage}[t]{0.48\linewidth}
\captionsetup{justification=raggedright,singlelinecheck=true}
\centering
\includegraphics[width=0.95\linewidth]{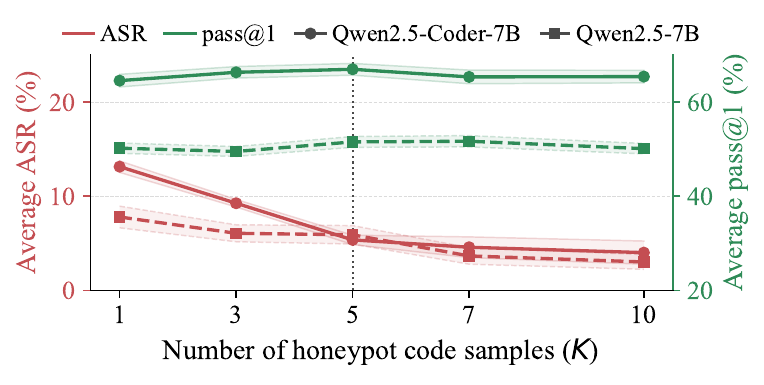}
\caption{Sensitivity of \defense to the number of honeypot code samples. Shaded regions indicate standard deviations across repeated runs.}
\label{fig:rq5-k}
\end{minipage}
\end{figure}

\bi{Results.}
Figure~\ref{fig:rq5-grammar} reports the results under different grammars, and Figure~\ref{fig:rq5-k} reports the results with different numbers of honeypot code samples.

\bi{Under our experimental setting, \attack improves attack effectiveness across all grammars.}
For example, on \qwencoderseven, the ASR remains far below 40\% without GCD, but becomes higher than 70\% once GCD is applied, regardless of which grammar is used.

\bi{As the number of honeypot code samples increases, safety improves while utility remains largely unchanged.}
As shown in Figure~\ref{fig:rq5-k}, when $K$ increases from 1 to 10, ASR shows a decreasing trend, whereas \pass{1} remains nearly unchanged.
This result supports our design choice of sampling multiple honeypot code responses for each malicious request during training.
A larger $K$ provides more preference pairs that compare harmful code with semantically harmless alternatives, helping the model choose honeypot code under GCD.

\summary{Answer to RQ5}{\attack remains effective across different grammars, while \defense benefits from using more honeypot code samples, improving safety without degrading benign utility.}

\section{Discussion}
\label{sec:8-discussion}

\FloatBarrier
\subsection{Can \defense Remain Robust under an Adaptive Attack?}
\label{sec:adaptive}

To examine whether \defense remains robust against stronger attackers, we further consider an adaptive attacker who is aware that \defense trains the model to generate honeypot code and can revise the grammar during the attack.
This setting is stricter than the main \attack evaluation, where the attacker simply applies a fixed off-the-shelf code grammar.
Specifically, for each malicious prompt $p$, the attacker uses \deepseekpro~\cite{deepseekv4} as an attack proxy to tighten the grammar for at most $R$ rounds.
In each round, the proxy observes the current response and proposes a revised grammar intended to exclude the observed honeypot behavior.
The target model is then queried again under the tightened grammar.
After $R$ rounds, we use the response from the final round as the attack result.

We conduct this adaptive evaluation on \qwencoderseven, \qwenseven, and \llamaeight, setting $R=10$.
As shown in Table~\ref{tab:adaptive}, the ASR and MR of \defense show little increase under the adaptive attack, and in some cases even decrease.
We attribute this robustness to the structural diversity of honeypot code learned by \defense.
Because the safe behavior is not anchored to a single identifiable syntactic structure, the adaptive attacker cannot easily suppress it through grammar tightening without also excluding many code structures needed to express malicious functionality.

\begin{table}[!htbp]
\centering
\caption{ASR (\%) and MR (\%) under \attack and the adaptive attack against models aligned with \defense. Lower values indicate better safety.}
\label{tab:adaptive}
\small
\setlength{\tabcolsep}{4.5pt}
\renewcommand{\arraystretch}{1.08}
\begin{adjustbox}{max width=\columnwidth,center}
\begin{tabular}{llcccc}
\toprule
\multirow{2}{*}{\textbf{Model}}
& \multirow{2}{*}{\textbf{Benchmark}}
& \multicolumn{2}{c}{\textbf{\attack}}
& \multicolumn{2}{c}{\textbf{Adaptive}} \\
\cmidrule(lr){3-4}
\cmidrule(lr){5-6}
& & \textbf{ASR} & \textbf{MR}
& \textbf{ASR} & \textbf{MR} \\
\midrule
\qwencoderseven & \dataname{RMCBench} & 7.69 & 4.40 & 6.04 & 0.16 \\
                & \dataname{MalwareBench} & 3.44 & 1.15 & 4.39 & 0.92 \\
\qwenseven      & \dataname{RMCBench} & 7.88 & 3.11 & 12.64 & 7.32 \\
                & \dataname{MalwareBench} & 3.33 & 0.62 & 8.61 & 1.09 \\
\llamaeight     & \dataname{RMCBench} & 6.04 & 3.66 & 10.80 & 2.38 \\
                & \dataname{MalwareBench} & 9.69 & 3.02 & 8.79 & 4.21 \\
\bottomrule
\end{tabular}
\end{adjustbox}
\end{table}

\FloatBarrier
\subsection{Reliability of LLM Judgment}
\label{sec:judge}

Following prior work~\cite{rmcbench, malwarebench, li2025goal}, we use an LLM judge to assess the safety of generated responses.
To validate its reliability, we manually evaluated 100 randomly sampled responses using the same criteria.
Human and LLM judgments reached agreement rates of 87\% for ASR and 85\% for MR, supporting the reliability of our evaluation.

\FloatBarrier
\subsection{Threats to Validity}
\label{sec:threat}

\ding{182} \bi{Coverage of GCD Settings.}
The effectiveness of \attack may vary across different GCD settings. Different inference engines and API providers may implement GCD in slightly different ways, which can affect the absolute attack success rate. To mitigate this threat, we evaluate \attack in both local and API-based deployment settings, and further test different code grammars in RQ5. Therefore, our results should be viewed as evidence of a general risk of GCD, rather than a claim that all GCD implementations behave identically.

\ding{183} \bi{Coverage of Evaluation Benchmarks.} 
Our evaluation may not cover all possible malicious code generation scenarios. To mitigate this threat, we use two complementary malicious-code benchmarks rather than relying on a single dataset. \dataname{RMCBench} covers 10 types of malicious coding scenarios, while \dataname{MalwareBench} covers 6 types of malware-related scenarios. This benchmark diversity reduces the risk that our conclusions are tied to one specific set of malicious requests. 

\ding{184} \bi{Ethical Considerations.}
In this work, we reveal that grammar-constrained decoding can be exploited to jailbreak leading LLMs and induce them to generate malicious code. This finding may raise concerns about potential misuse. To mitigate this threat, we release the source code of \attack only to researchers for controlled research use, while fully releasing the source code of \defense. We also provide \defense as a mitigation approach and encourage the community to adopt it to enhance the intrinsic safety of LLMs.

\section{Conclusion}
\label{sec:9-conclusion}

In this paper, we propose \attack and \defense. Through comprehensive experiments on 10 popular LLMs across 4 benchmarks, we show that \attack can effectively bypass the existing safety alignment of both locally deployed and API-based LLMs. We further demonstrate that \defense can effectively restore model safety under GCD while preserving benign utility. We hope this work draws greater attention to the potential security implications of GCD and inspires further efforts toward building safer intelligent systems.

\section{Data Availability}

The source code, related datasets, and supplementary material can be accessed at: \url{https://github.com/THU-Agent/CodeSpear-CodeShield}.

\section*{Acknowledgment}

This work was supported by the National Natural Science Foundation of China (No. 62602364), the Beijing Natural Science Foundation (No. 4264107), and the Tsinghua University–Keystone Electrical (Zhejiang) Co., Ltd. Joint Research Center for Embodied Multimodal Artificial Intelligence (JCEMAI).

\bibliographystyle{preprint}
\bibliography{ref}

\end{document}